# Experimental and Theoretical Investigations of Stochastic Oscillatory Phenomena in a Nonrelativistic Electron Beam with a Virtual Cathode

Yu. A. Kalinin, A. A. Koronovskiĭ, A. E. Hramov, E. N. Egorov, and R. A. Filatov

*Chernyshevsky Saratov State University, Astrakhanskaya ul. 83, Saratov, 410026 Russia*

**Abstract**—Results are presented from experimental investigations of oscillatory phenomena in an electron beam with a virtual cathode in a diode gap with a decelerating field. Experiments have revealed a stochastic broadband generation of the microwave oscillations of a virtual cathode in a decelerating field. Numerical simulations based on a simple one-dimensional model have shown that the onset of the stochastic generation and the broadening of the oscillation spectrum with increasing beam deceleration rate are governed by the processes of regrouping of the electrons in a beam with a virtual cathode.

## 1. INTRODUCTION

A new class of devices of high-power microwave electronics—one in which an electron beam with a virtual cathode (VC) is used as an active medium—was proposed in the late 1970s and early 1980s [1–4] and is still a subject of active research [5–8]. Even the first experiments and numerical simulations revealed the complicated unsteady dynamics of radiation from vircators, i.e., microwave oscillators with a VC [9–13]. Quite a number of theoretical and experimental papers were devoted to studying the stochastic generation of radiation in vircators [5–7, 13–17], the processes responsible for the onset of coherent spatiotemporal structures in an electron beam with a VC and for their interaction [16, 18–21], and the mutual synchronization of VC-based devices [6, 22–25]. In particular, by applying different methods for separating out coherent structures, it was shown theoretically [18–20, 26] that the complication of the spectral composition of radiation from a vircator is closely related to the formation of electron structures in a beam with an overcritical current. However, no detailed experimental investigations of the stochastic generation of radiation in VC-based systems, as well as of the physical mechanisms governing the complicated behavior of an electron beam with a VC, were carried out, largely because of the difficulties in diagnosing relativistic electron beams (REBs).

One of the simplest systems for studying the formation of a VC and its unsteady stochastic oscillations is a Pierce diode [17, 27–29]—a system of two infinitely wide parallel equipotential grids penetrated by an electron beam with an overcritical current. A system like this, in which the electron motion is thought of as being one-dimensional, can be regarded as a model of the drift space and can also serve to investigate the nonlinear dynamics of the processes in such high-power electronic devices as vircators [29]. In recent years, a model of this type has also attracted attention as one of the simplest models of a distributed self-oscillating electron–plasma system for studying stochastic oscillations [17, 26, 29–31].

Note that experimental investigation and practical development of VC-based oscillators is a rather difficult task because of the necessity of using intense REBs with overcritical currents, i.e., with currents above the limiting vacuum current [32]. This is why a detailed study of the parameters of the generation of radiation in vircators, as well as of the physical processes occurring in an electron beam with a VC, is impossible. The conditions of experiments on microwave generation by a VC can be "relaxed," e.g., by using systems with additional electron deceleration; in this case, an unsteady VC forms at comparatively low beam currents in a decelerating field with a strength above a certain critical value.

In this context, it is of great interest to investigate a modified Pierce diode in the form of a drift gap in which an unsteady oscillating VC forms at the expense of strong beam deceleration (an electron beam with an overcritical perveance). In such a system, a VC can form and a stochastic broadband signal can be generated at low electron beam currents and densities, which makes possible a detailed experimental study of the physical processes occurring in a beam with a VC with the help of experimental methods used in microwave electronics [33]. Note that this type of system with a decelerated electron beam has not yet been studied theoretically or experimentally.

The objective of the present paper is to experimentally investigate the oscillatory phenomena in a nonrelativistic electron beam formed by an electron optical system in a diode gap with a decelerating potential and to analyze the behavior of an electron beam in a drift gap with a decelerating field by means of numerical simulations.

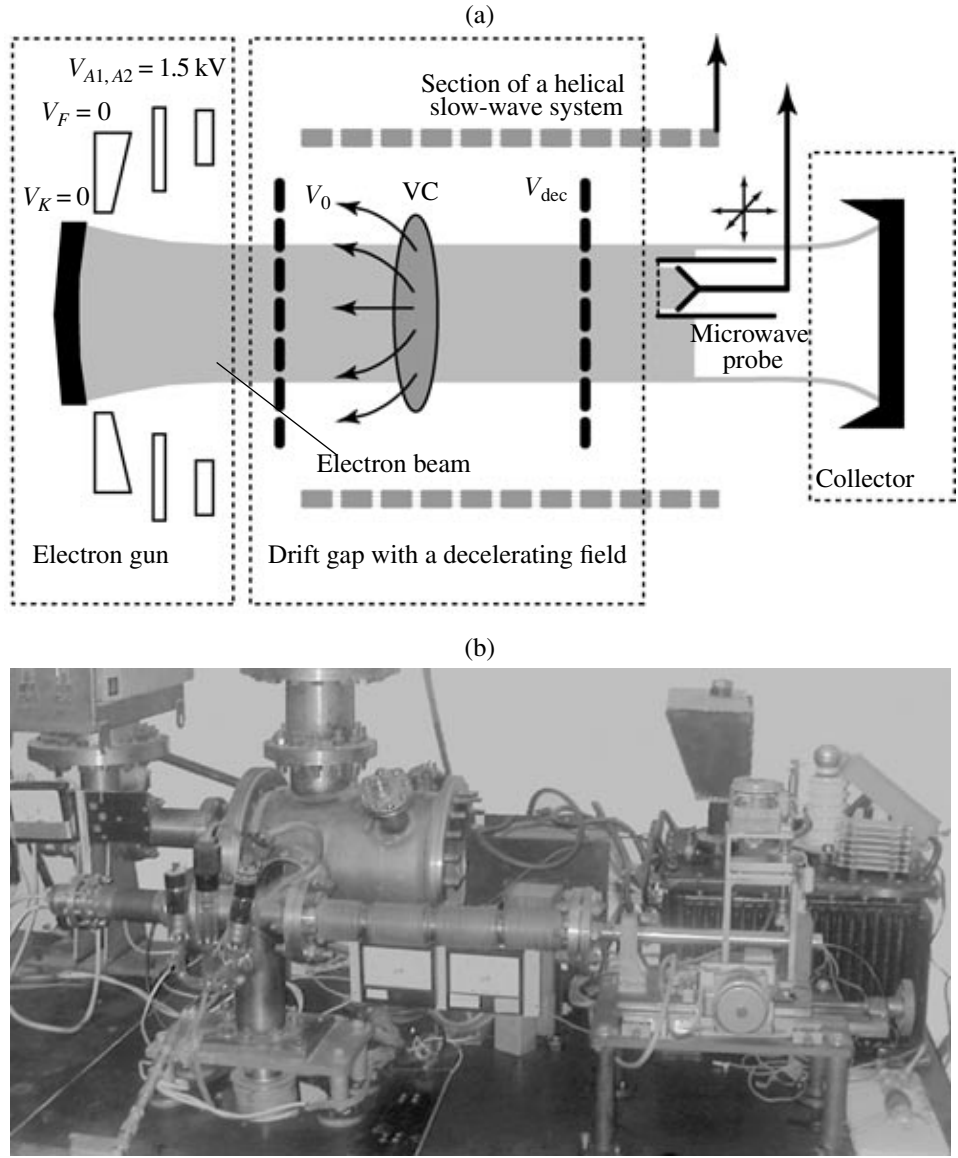

**Fig. 1.** (a) Schematic and (b) external view of the experimental device for investigating stochastic oscillations in an electron beam with a VC in a system with an overcritical perveance.

## 2. EXPERIMENTAL TECHNIQUES

As an object for experimental investigations of oscillations in an electron beam with a VC, we chose a diode scheme in which a beam formed by an electron optical system (EOS) was injected into a diode consisting of two grid electrodes (Pierce diode) with a decelerating field (a schematic of the experimental setup is shown in Fig. 1a). The decelerating field was produced by applying a negative (with respect to the first (entrance) grid) potential $V_{dec}$ to the second (exit) grid.

The electron source was a hot cathode whose current was limited by the space charge. The EOS formed an axisymmetric cylindrical converging electron beam in the drift space, which was surrounded by a solenoid creating a guiding longitudinal magnetic field. The magnetic field was varied from 0 to 250 G; the voltage used to accelerate an electron beam was equal to 1.5 kV; and the beam current at the exit from the EOS was varied over a range of 50–100 mA, depending on the cathode heating voltage and on the accelerating potential. The spread in electron velocities at the exit from the EOS was very small (less than 0.2%), so the electron beam could be considered monoenergetic.

The electron beam formed by the EOS entered the region between the grids (the diode gap). The potential

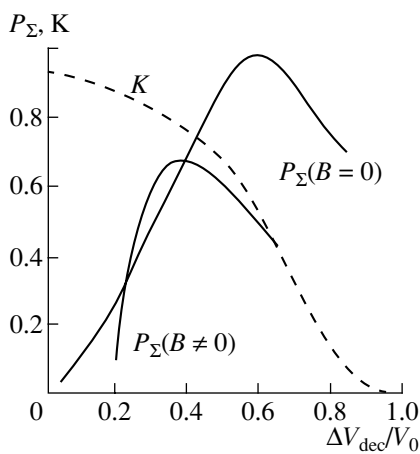

**Fig. 2.** Global interaction power $P_\Sigma$ (for different magnetic field strengths) and the coefficient $K$ of current transmission through the diode (at a magnetic field strength of $B = 220$ G) as functions of the normalized potential difference $\Delta V_{dec}/V_0$ between the grids.

$V_0$ of the first grid was equal to the anode potential $V_A$ (the accelerating voltage), and the potential of the second grid, $V_{dec} = V_0 - \Delta V_{dec}$, was varied from values such that $V_{dec}/V_0 = 1$ ($\Delta V_{dec} = 0$, the electron beam was not decelerated at all) to $V_{dec}/V_0 = 0$ ($\Delta V_{dec} = V_0$, the electron beam was decelerated to a complete stop). The quantity $\Delta V_{dec}$ has the meaning of the potential difference between the grids: it produces the decelerating field in the diode gap. Note that the operating regime with a zero decelerating potential, $\Delta V_{dec} = 0$, corresponds to that of a "classical" vircator, and the operating regime with $\Delta V_{dec}/V_0 = 1$ is that of a reflex triode [6, 34].

An increase in the decelerating potential of the second grid can be considered as an increase in the perveance of the electron beam in the diode gap:

$$p = I/\overline{V}^{3/2}, \qquad (1)$$

where $\overline{V} = (V_0 + V_{dec})/2$ is the effective potential.

As the decelerating potential difference $\Delta V_{dec}$ between the grids of the diode gap was increased, a VC was observed to form in the system at a certain critical value of the beam perveance, $p = p_{cr}$. The electron beam was modulated by the temporal and spatial oscillations of the VC; moreover, some of the electrons were reflected from the VC back toward the entrance grid. As a result, the VC was observed to execute stochastic oscillations, whose shape and power depended substantially on the potential difference $\Delta V_{dec}$ between the grids of the diode gap.

The noisy oscillations in an electron beam were analyzed using a broadband section of a helical slow-wave system (HSS) loaded on an absorbing insertion and on an energy extraction unit, and also using a section of a broadband coaxial line loaded on a resistance—an RF probe that could be moved in three mutually perpendicular directions [33]. The coaxial probe was equipped with a decelerating grid in order to analyze the distribution of charged particles over longitudinal velocities. Within the diode gap, a velocity- and density-modulated electron beam excited the broadband section of the HSS, the signals from which were processed by an SCh-60 spectrum analyzer. This made it possible to determine the spectral power density of the noisy oscillations generated by an electron beam with a VC.

Experimental investigations were carried out using a sectional vacuum device with continuous pumping-out (the minimum residual gas pressure being about $10^{-6}$–$10^{-7}$ torr). A photograph of the experimental device is presented in Fig. 1b.

## 3. EXPERIMENTAL RESULTS

Oscillations in the system under examination are governed by the presence of a VC in the diode gap with a decelerating field. This is why it is very important to study how the processes occurring in an electron beam with a VC depend on the decelerating potential of the second grid. The generation processes in a system with a VC are also highly sensitive to the strength of the external focusing magnetic field, [35, 36] (this question, too, will be discussed in the present section).

Figure 2 shows the normalized global oscillation power as a function of the normalized potential difference $\Delta V_{dec}/V_0$ between the grids in the presence and in the absence of a magnetic field ($B = 220$ and $0$ G, respectively). We can see that there is an optimum value of the decelerating potential at which the global power $P_\Sigma$ of oscillations in a beam with a VC is maximum. Analyzing Fig. 2, we also note that the oscillation power depends strongly on the external focusing magnetic field. The power of the oscillations generated by the VC in the absence of a magnetic field ($B = 0$) is substantially higher and is maximum at a decelerating potential of $\Delta V_{dec}/V_0 \sim 0.5$. In the presence of the focusing magnetic field, the global power of the oscillations generated by the VC is lower. The stronger the magnetic field, the lower the potential difference $\Delta V_{dec}$ between the grids at which the global oscillation power in the beam is maximum.

Figure 2 also shows the coefficient $K$ of current transmission through the diode gap as a function of the potential difference between the grids in the presence of the external magnetic field ($B = 220$ G). The current transmission coefficient is defined as the ratio of the current that has passed through the second grid, $I_{out}$, to the current $I_0$ of the beam formed by the electron gun: $K = I_{out}/I_0$. It can be seen that, in contrast to the global oscillation power $P_\Sigma$, the current transmission coefficient $K$ decreases as the decelerating potential increases. This indicates that an increase in the decelerating potential difference $\Delta V_{dec}$ is accompanied by an increase in the number of electrons reflected from the VC (or from the diode region behind the VC) back toward the entrance grid.[1] At a certain, sufficiently high, potential difference $\Delta V_{dec}$ between the grids (whose magnitude depends weakly the strength of the external focusing magnetic field), almost all of the beam electrons are seen to be reflected from the VC, so the current transmission coefficient becomes nearly zero, $K \approx 0$. This is the case only when the additional electron deceleration is sufficiently strong, i.e., when the decelerating potential difference is comparable to the accelerating voltage, $\Delta V_{dec}/V_0 \approx 1$. In this case, almost all of the beam electrons that have passed through the potential barrier of an unsteady VC are turned by the high decelerating potential $V_{dec}$ at the second grid back to the injection plane, so the current transmission coefficient becomes rather small, $K \approx 0$. As the decelerating potential difference increases, the VC tends to be displaced toward the injection plane.

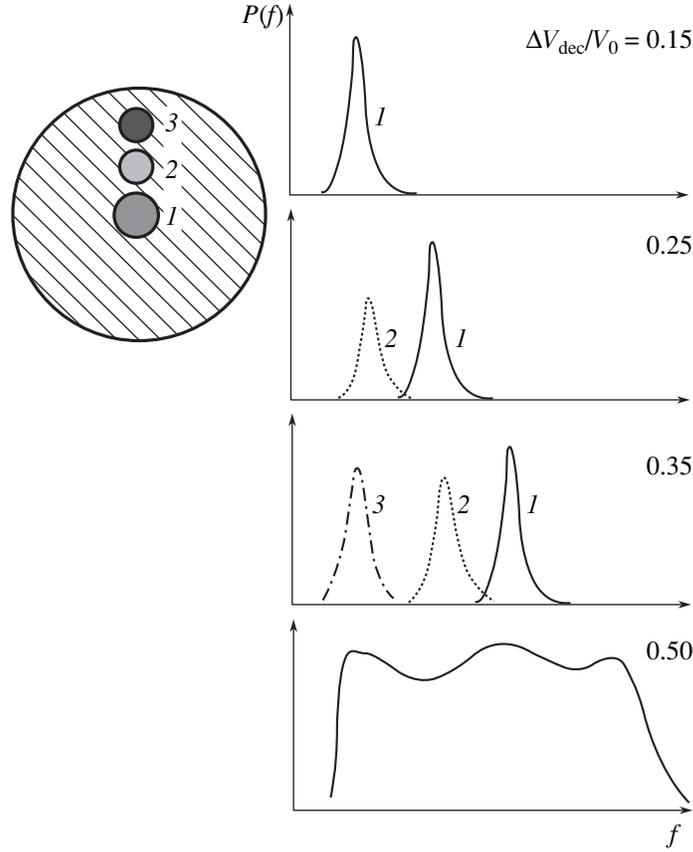

**Fig. 3.** Qualitative shapes of the spectra of the beam current oscillations recorded by an RF probe at different radial positions within the beam (shown schematically on the left) for different values of the decelerating potential difference.

Let us now consider how the generation process in an electron beam in the diode gap develops when the decelerating potential difference (i.e., the perveance $p$ of the beam in the space between the grids) is increased.

---
[1] Recall that an increase in the decelerating potential $\Delta V_{dec}$ corresponds to a decrease in the potential $V_{dec}$ of the second grid, so the electron beam is decelerated to a complete stop at $V_{dec} = 0$.

To do this, we analyze the spectra of the high-frequency current oscillations recorded by an RF probe at different radii $r_b$ within the beam at several values of the decelerating potential difference, each next being larger than the previous one. The results of the analysis of the development of the generation process are illustrated in Fig. 3, where we schematically show the radial positions within the beam at which high-frequency current oscillations were recorded and for which the power spectra were obtained with the help of the spectrum analyzer. An analysis of the results presented in Fig. 3 allows us to draw the following conclusions.

At small values of the decelerating potential difference between the grids in the diode gap, i.e., when the perveance $p$ of the electron beam is lower than a critical value $p_{cr}$, no beam oscillations were recorded. As the decelerating potential difference (i.e., the ratio $\Delta V_{dec}/V_0$) increases, the beam perveance reaches a certain critical value $p_{cr}$ at which a VC forms in the system. The VC begins to reflect some electrons back toward the first grid of the diode, so the coefficient $K$ of the transmission of the electron beam through the diode gap begins to decrease. The most important effect here is that the high-frequency current oscillations are recorded only at the center of the beam (see Fig. 3, point *1* and curve *1*, which refer to the case $\Delta V_{dec}/V_0 =$ 0.15). At larger radii in the beam cross section, no high-frequency current oscillations were observed; consequently, at a low deceleration rate, i.e., for a low beam perveance, $p \gtrsim p_{cr}$), a VC first arises at the center of the electron beam. As the beam deceleration rate becomes faster, the frequency of oscillations of the beam current at the beam center increases (cf. the positions of curve *1* for $\Delta V_{dec}/V_0 = 0.15$ and 0.25 in Fig. 3) and the beam current begins to oscillate at larger beam radii. An analysis of the power spectra for $\Delta V_{dec}/V_0 = 0.25$ and 0.35 shows that, as the beam deceleration rate increases, the beam current begins to oscillate first at radial position *2* and then at radial position *3* in the cross section of the electron beam. In this case, the high-frequency current oscillations at the periphery of the beam (i.e., in the beam's external layers) become more intense as the decelerating potential $\Delta V_{dec}$ (and, consequently, the electron beam perveance) increases. This indicates that, within the diode gap, the VC is distributed in both space (in the transverse direction) and time. As the decelerating potential increases, an unsteady electron-reflecting VC forms first at the center of the electron beam and then begins to expand radially. Following [8, 37], this can be interpreted as the formation of a VC in an electron beam that has the shape of a cup in the radial direction and is convex toward the injection plane. In terms of this radial structure of the VC, it is possible to explain why the oscillation frequency is different at different beam radii. Specifically, the larger the radius, the greater the longitudinal dimension of the potential well in the electron beam and, consequently, the lower the frequency of oscillations of the VC.

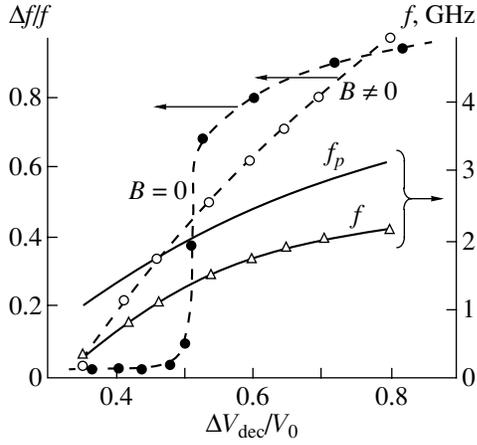

**Fig. 4.** Characteristic frequency and bandwidth of stochastic oscillations in a beam with a VC as functions of the potential difference $\Delta V_{dec}/V_0$ between the grids. Theoretical dependence (2) of the plasma frequency $f_p$ of the electron beam in the space between the grids of the diode gap is also shown.

A further increase in the electron deceleration rate was accompanied by the onset of broadband noisy oscillations whose spectrum was found to be essentially the same at different radial positions (i.e., over the entire beam cross section). A representative spectrum of the beam current oscillations for $\Delta V_{dec}/V_0 = 0.5$ is illustrated qualitatively in Fig. 3. In this case, the electrons in the beam execute oscillations over its entire cross section. After being reflected by the VC, the electrons can occur at arbitrary radii of the beam, giving rise to developed stochastic oscillations. A fundamentally important point here is that a description of this operating regime requires at least a two-dimensional theory of the dynamics of an electron beam in the VC region.

In view of such complications in the oscillation spectrum of a beam with a VC as the electron deceleration rate (i.e., the perveance of the beam in the diode gap) is increased, it is necessary to consider the parameters of these spectra in more detail. Figure 4 shows the normalized frequency bandwidths $\Delta f/f$ of the generated oscillations that were measured from the oscillation power spectra recorded by a portion of a helix in the VC region at different strengths of the external magnetic field. Figure 4 also shows the characteristic generation frequency (i.e., the frequency at which the power in the spectrum is generated most intensely) in the system as a function of the decelerating potential difference $\Delta V_{dec}/V_0$. The frequency bandwidth $\Delta f$ was determined at a level of 3 dB in the oscillation power spectrum.

From Fig. 4 it follows that the characteristic generation frequency in a VC-based system increases monotonically with increasing decelerating potential difference $\Delta V_{dec}/V_0$ (or equivalently beam perveance). In [38, 39], it was shown theoretically that the characteristic frequency of oscillations of the VC is proportional to the plasma frequency of the electron beam, $f_{VC} \propto f_p$. In addition, Fig. 4 shows the theoretical dependence of the reduced plasma frequency $f_p$ of an electron beam on the potential difference between the grids of the diode gap:

$$f_p^2 = \frac{2^{3/2}\pi\sqrt{\eta}}{\varepsilon_0}\frac{p\bar{V}}{r^2}, \qquad (2)$$

where $p$ and $r$ are the perveance of the electron beam and its radius, $\bar{V} = (V_0 + V_{dec})/2$, $\eta = \dfrac{e}{m_e}$ is the electron charge-to-mass ratio, and $\varepsilon_0$ is the dielectric constant. A comparison between the characteristic generation frequencies shows that the frequency of oscillations of the VC is related to the plasma frequency by

$$f_{VC} = kf_p, \qquad (3)$$

where $k \approx 0.5$ is a numerical factor. Note that, in [13, 38], when investigating the dependence of the frequency of oscillations of a VC on the plasma frequency in systems without additional electron deceleration, the numerical factor was found to be equal to $k \approx 2$. It can be suggested that the discrepancy between the values of the numerical factors $k$ that were obtained empirically in the experiment and were found in [13, 38] can be attributed to the additional deceleration of the electron beam. Note also that, according to formula (2), the dependence of the frequency $f_{VC}$ on the decelerating potential is described by the relationship $f_{VC} \propto p\sqrt{\overline{V}}$, which agrees well with experiment.

The normalized frequency bandwidth $\Delta f/f$ of the oscillations generated in a beam with a VC also increases as decelerating potential difference in the diode gap increases. However, the dependence $\Delta f/f(V_{dec})$ at a strong focusing magnetic field ($B = 220$ G) differs from that in the absence of a magnetic field ($B = 0$). In the first case, the generation frequency bandwidth increases monotonically with the decelerating potential difference $\Delta V_{dec}$. When the magnetic field is absent, the generation frequency bandwidth increases in a jumplike manner at $\Delta V_{dec}/V_0 \approx 0.3$. A further increase in the decelerating potential of the second grid of the diode gap (i.e., in the perveance of the electron beam) is accompanied by a monotonic increase in the frequency bandwidth $\Delta f/f$.

Hence, we can speak of two characteristic operating regimes of the electron–wave VC-based oscillator under consideration: the regime of the generation of nearly regular narrowband oscillations, which occurs when the beam perveance is slightly above the critical value (i.e., when the deceleration rate of the beam in the diode gap is slow), and the regime of the generation of stochastic oscillations with a broadband spectrum, which occurs when the electron beam perveance is high (i.e., when the beam deceleration is strong).

An important property of the VC-based oscillator under consideration is its ability to generate microwave radiation in the absence of an external magnetic field. The operating regime without a magnetic field is characterized by the generation of a stochastic broadband signal with a slightly irregular spectrum within the electron beam. This circumstance can be very useful in creating sources of broadband noisy signals with more than an octave bandwidth. Therefore, it is important to analyze how the global power of oscillations in an electron beam with a VC depends on the strength of the external magnetic field. The above experimental results (see Fig. 2) show that the maximum global oscillation power increases with decreasing magnetic field.

Figure 5 shows the normalized global power $P_\Sigma$ of oscillations in a beam with a VC as a function of the magnetic field in the regimes of the narrowband generation (the normalized decelerating potential of the second grid is equal to $\Delta V_{dec}/V_0 = 0.2$) and of the stochastic broadband generation (for $\Delta V_{dec}/V_0 = 0.5$). We can see that, in the regime of the weakly stochastic narrowband generation, the global oscillation power $P_\Sigma$ is maximum at a certain magnetic field strength, whereas, in the broadband generation regime, the power $P_\Sigma$ is maximum in the absence of a magnetic field ($B = 0$).

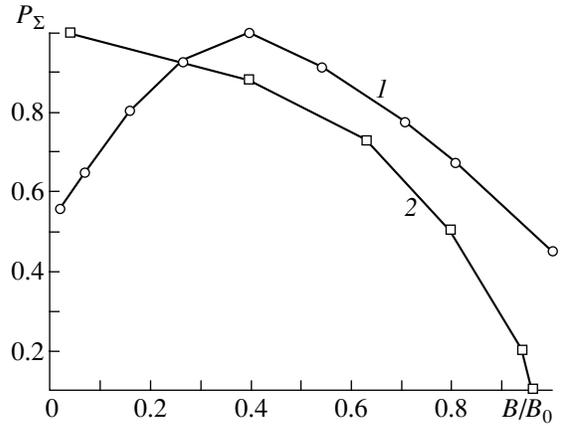

**Fig. 5.** Global powers of (*1*) narrowband and (*2*) broadband oscillations in an electron beam as functions of the external magnetic field ($B_0 = 250$ G).

Let us now consider distinctive features of the physical processes in an electron beam with a VC. An important characteristic that makes it possible to analyze the state of the beam in the VC region and which can be obtained experimentally is the velocity distribution of the beam electrons that have passed through the VC and have reached the second grid.

Figure 6 presents the velocity distribution functions of the electrons behind the second grid for different values of $\Delta V_{dec}/V_0$. The distributions were recorded by an RF probe placed at the exit from the drift space. Figure 7 displays the mean electron velocities $v_{mean}$ and the normalized velocity spread $\Delta v/v_{mean}$ calculated for several values of the potential difference $\Delta V_{dec}/V_0$ from the experimentally measured electron velocity distribution functions. We can see that the shape of the electron distribution function changes radically as the oscillations of the VC grow with increasing deceleration rate of the electron beam in the diode gap. The most important effects here are the broadening of the velocity distribution function $\Delta v/v_{mean}$ and the increase in its irregularity as the beam deceleration rate increases and, consequently, the stochastic oscillations in a beam with a VC become more complicated. Such behavior of the electron velocity distribution with increasing beam perveance can be attributed to the formation of several electron structures (bunches) in the VC region [16]. For instance, when the beam deceleration rate is slow, i.e., when the oscillations in the system are nearly regular, the electron distribution function is close to a δ function. This indicates that the only structure that forms in the system is a VC. As the beam deceleration rate increases, the electron velocity distribution becomes rather complicated (acquires a multipeak structure); this points to the regrouping of electrons and, consequently, the formation of electron structures (bunches) in the VC region.

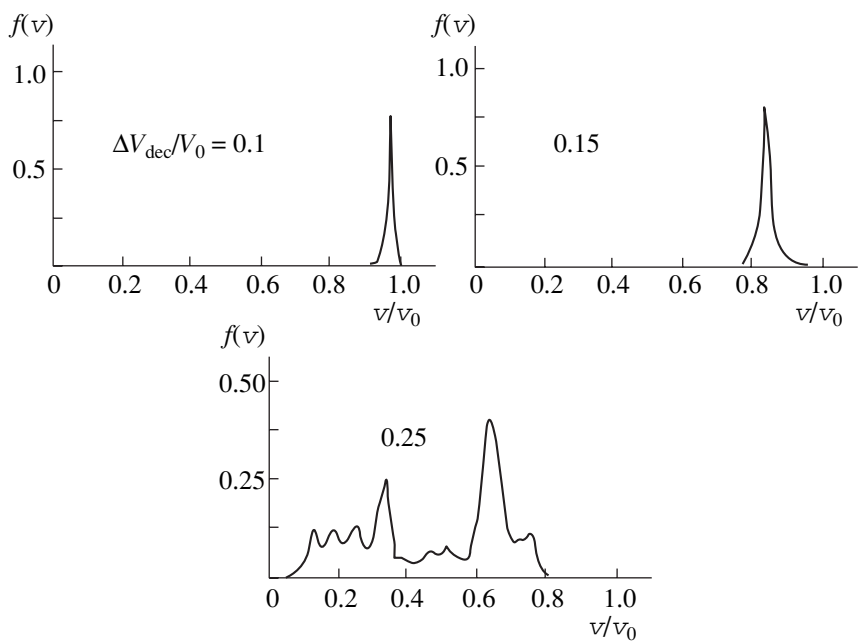

**Fig. 6.** Electron velocity distribution function at the exit from the second grid for different values of the decelerating potential $\Delta V_{dec}/V_0$.

Finally, we consider the experimentally measured parameters of microwave generation by an electron beam with a VC when there are positively charged ions in the drift space. The presence of positive ions (which corresponds to a higher pressure of the residual gas), on the one hand, decreases the electron density and partially neutralizes the space charge of the electron beam, thereby preventing the formation of a VC [29], and, on the other hand, gives rise to additional ion oscillations (such as relaxation, plasma, and radial oscillations) [40]. This, according to the experiment, makes the power spectra of the noisy oscillations in such beams more regular. Figure 8 shows how the modulation depth $N$ (defined as $N = P_{max}/P_{min}$, where $P_{max}$ and $P_{min}$ are the maximum and minimum spectral powers in the working frequency band) of the oscillation spectrum of an electron beam with a VC depends on the pressure of the residual gases in the interaction space of the oscillator. The potential of the second grid was equal to $\Delta V_{dec}/V_0 = 0.5$. It is clear from Fig. 8 that, at a pressure of $P \sim 5 \times 10^{-7}$ torr, the modulation depth is about 7–8 dB, whereas at a pressure of $P \sim 10^{-5}$ torr, it is about 3 dB. We thus can see that an increase in the pressure of the residual gases in the drift space of an electron–wave VC-based oscillator leads to an improvement in the spectral parameters of the noisy broadband generation: the continuous spectrum of the generated noisy broadband radiation becomes more regular and, accordingly, more uniform.

## 4. NUMERICAL SIMULATIONS OF THE OSCILLATIONS OF A VIRTUAL CATHODE IN A PIERCE DIODE WITH A DECELERATING FIELD

Here, we consider the results from numerical simulations of unsteady processes by means of a simple model of an electron beam with a VC. We also compare the theoretical and experimental results on the oscillations of the VC.

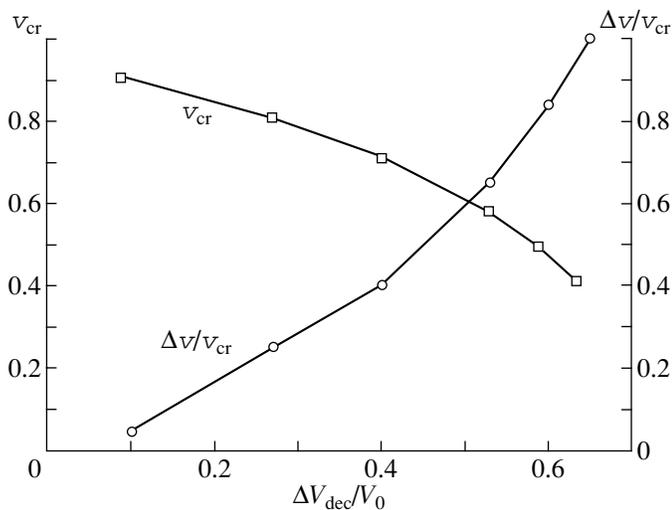

**Fig. 7.** Mean velocity $v_{mean}$ of the electrons that have passed through the second grid and the spread in its values, $\Delta v/v$, for different values of the decelerating potential difference $\Delta V_{dec}/V_0$ between the grids of the diode gap.

The unsteady nonlinear processes occurring in a charged particle beam with a VC were simulated in terms of a one-dimensional model of a drift gap with a decelerating field by the particle-in-cell (PIC) method [41, 42]. Of course, the assumption of one-dimensional motion of the electron beam is not valid for all operating regimes of the experimental electron–wave VC-based oscillator under consideration. It might be supposed, however, that the main physical mechanisms for the formation of a VC in a diode gap with a decelerating field in the one-dimensional case would be qualitatively similar to those in a more complicated case of two-dimensional motion of an electron beam with a VC.

Let us briefly outline the scheme for numerical simulations. In plane geometry, the electron beam is modeled as a system of macroparticles (charged plane sheets) injected into the interaction space with a constant velocity at equal time intervals. We switch from the dimensional parameters, namely, the potential $\varphi$, the space charge field $E$, the electron density $\rho$, the electron velocity $v$, the spatial coordinate $x$, and the time $t$, to the following dimensionless variables, which are marked by a prime:

$$\varphi = (v_0^2/\eta)\varphi', \quad E = (v_0^2/L\eta)E', \quad \rho = \rho_0\rho', \quad v = v_0 v', \quad x = Lx', \quad t = (L/v_0)t', \quad (4)$$

where $\eta = e/m_e$, $v_0$ and $\rho_0$ are the static (unperturbed) electron beam velocity and electron density, and $L$ is the length of the drift gap. In what follows, the primes by the dimensionless variables will be omitted.

For each of the charged sheets (macroparticles), we solve the dimensionless nonrelativistic equations of motion

$$\frac{d^2 x_i}{dt^2} = -E(x_i), \quad (5)$$

where $x_i$ is the coordinate of the $i$th sheet and $E(x_i)$ is the space charge field at the point with the coordinate $x_i$.

In order to calculate the strength of the space charge field and its potential and also the charge density, we introduce a uniform spatial mesh with a spacing $\Delta x$. In the quasistatic approximation, the potential of the space charge field is determined from the one-dimensional Poisson's equation

$$\frac{\partial^2 \varphi}{\partial x^2} = \alpha^2 \rho(x), \quad (6)$$

where $\alpha = \omega_p L/v_0$ is the so-called Pierce parameter [27]. In Eq. (6), the density of the positive ion background is assumed to be low enough to be ignored in simulating the oscillatory processes in an electron beam. The space charge field $E$ is calculated by numerically integrating the values of the potential that have been obtained from Poisson's equation.

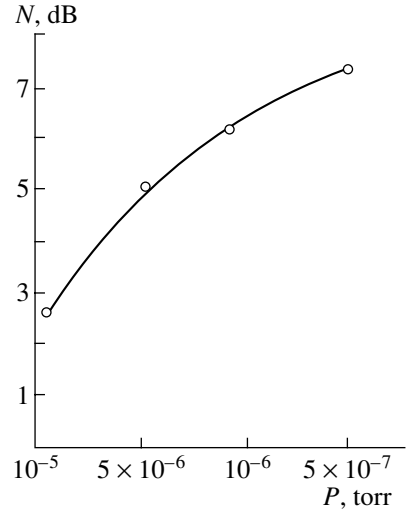

**Fig. 8.** Dependence of the modulation depth $N$ of the oscillation spectrum on the residual gas pressure in the diode gap.

Poisson's equation (6) is supplemented with the boundary conditions

$$\varphi(x=0) = \varphi_0, \quad \varphi(x=1) = \varphi_0 - \Delta\varphi, \quad (7)$$

where $\varphi_0$ is the accelerating potential (in the dimensionless units used here, it is equal to $\varphi_0 = 1$) and $\Delta\varphi$ is the decelerating potential difference between the grids.

The space charge density is calculated using the PIC method, i.e., by linearly weighing the contributions of macroparticles (charged sheets) to its mesh values—a technique that reduces the noise introduced in computations by the mesh [42]. In this method, the space charge density in the $j$th mesh point, i.e., at the point with the coordinate $x_j = j\Delta x$, is expressed as

$$\rho(x_j) = \frac{1}{n_0}\sum_{i=1}^{N}\Theta(x_i - x_j). \quad (8)$$

Here, $x_i$ is the coordinate of the $i$th macroparticle; $N$ is the total number of macroparticles; the parameter $n_0$ of the computational scheme is equal to the number of macroparticles in a cell in an unperturbed state; and

$$\Theta(x) = \begin{cases} 1 - |x|/\Delta x, & |x| < \Delta x, \\ 0, & |x| > \Delta x \end{cases} \quad (9)$$

is a piecewise-linear form function, which determines the procedure of weighting the contribution of a macroparticle on a spatial mesh with the spacing $\Delta x$.

The macroparticles that have reached the exit grid, as well as those that have been reflected from the VC and have reached the first (entrance) grid, are absorbed at the boundaries. This, in fact, corresponds to the reditron model of a VC-based oscillator [6, 43]. It should be noted, however, that the use of the reditron model in numerical simulations brings about simplifying assumptions because, in a model experimental device, the electrons can repeatedly pass from the accelerating gap to the deceleration region.

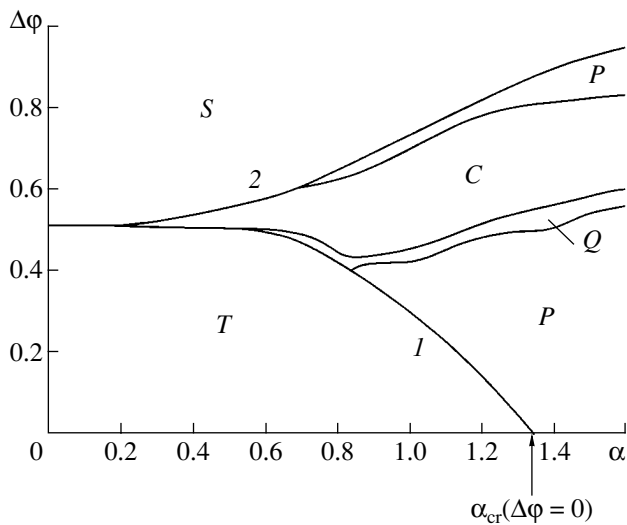

**Fig. 9.** Characteristic regimes of the behavior of an electron beam inside a diode gap with a decelerating field in the "Pierce parameter–decelerating potential difference" plane: (*C*) stochastic oscillations in a beam with a VC, (*Q*) quasi-periodic (two-frequency) oscillations of the VC, (*P*) periodic oscillations of the VC, (*S*) steady VC, and (*T*) total transmission of the electron beam through the diode gap. The arrow shows the parameter value $\alpha_{cr} = 4/3$ at which an unsteady VC forms in a system without a decelerating field, curve *1* is the boundary at which the steady single-stream state loses its stability (i.e., an unsteady VC forms), and curve *2* is the boundary at which a steady VC forms in a beam that reflects all the beam electrons.

The main parameters of the numerical scheme—the number $N_C$ of points in the spatial mesh and the number $n_0$ of macroparticles in a cell in an unperturbed state—were chosen to be $N_C = 800$ and $n_0 = 24$ (which correspond to $N = 19\,200$ macroparticles in the computation region in an unperturbed state). These values of the parameters of the numerical scheme were chosen so as to provide the required accuracy and reliability of calculations aimed at analyzing complicated nonlinear processes (including deterministic chaos) in the electron-plasma system under investigation [42, 44]. The equations of motion were solved by a leap-frog scheme of second-order accuracy [42], and Poisson's equation was integrated by the error-correction method [45].

Figure 9 illustrates the characteristic regimes of oscillations of the electron beam in the diode gap in the plane of the parameters $\alpha$ and $\Delta\varphi$, which are, respectively, the Pierce parameter and the decelerating potential difference between the grids. Let us discuss these regimes in more detail, by referring to Figs. 9 and 10, if necessary. Figure 10 presents the oscillation power spectra $P(f)$ and the phase portraits (obtained by the Takens method [46]) of the electric field oscillations $E(x = 0)$ at the first (entrance) grid of the diode gap for a Pierce parameter equal to $\alpha = 0.9$ and for different values of the decelerating potential difference $\Delta\varphi$.

Domain *T* in Fig. 9 refers to the regime in which the electron beam is fully transmitted through the drift gap; in this case, no VC forms in the electron beam and no oscillations are excited there. Curve *1* corresponds to the critical values of the control parameters (the Pierce parameter $\alpha$ and potential difference $\Delta\varphi$) at which the system becomes unstable and an unsteady VC, oscillating in both space and time, arises in the beam. When the beam is not decelerated ($\Delta\varphi = 0$), the critical value of the Pierce parameter at which an unsteady VC forms in the system is equal to $\alpha_{cr} = 4/3$ [29] (in Fig. 9, this critical value $\alpha_{cr}$ at the abscissa is indicated by the arrow). As the decelerating potential difference $\Delta\varphi$ increases, the boundary of the region in which an unsteady VC can appear in the parameter space is displaced toward smaller values of the Pierce parameter. This makes it possible, by increasing the decelerating potential at the second grid of the diode, to achieve microwave generation by a VC at beam currents lower than those in a system without electron deceleration (a "classical" scheme of a VC-based oscillator [6]).

Curve *2* in Fig. 9 corresponds to the values of the control parameters at which the oscillations in the system are suppressed and a steady VC forms in the beam that reflects all the electrons that were injected into the diode gap. This regime, which is denoted by the symbol *S* in Fig. 9, corresponds to the strong deceleration of the electron beam; it can be described analytically in terms of the steady-state theory of electron beams with an overcritical current (see, e.g., [47]).

Domains *P*, *Q*, and *C* in Fig. 9 correspond to different characteristic oscillation regimes of an electron beam with a VC. Analyzing the pattern of the regimes, we can conclude that, as the decelerating potential in the system increases, an electron beam with a VC sequentially evolves through different oscillations regimes. Let us examine these regimes in more detail.

Two domains *P* correspond to the regimes of regular oscillations of the VC. In the lower of these two regions, regular oscillations in a system with a decelerated electron beam occur at low decelerating potentials of the second grid (i.e., when the beam perveance $p$ only slightly exceeds the critical value $p_{cr}$ at which a VC appears in the system). For this regime with a slow deceleration rate of the electron beam, the parameters of the electric field oscillations are shown in Fig. 10a. In the upper region, regular oscillations occur at high deceleration rates of the electron beam (i.e., in Fig. 9, they occur near curve *2*, corresponding to a transition of the system to a state with a steady VC). The parameters of the regular oscillations for this case are shown in Fig. 10c, which was obtained for a high decelerating potential difference, $\Delta\varphi = 0.625$, between the grids of the diode gap. A comparison between Figs. 10a and 10c leads to the conclusion that, qualitatively, the shape of the oscillations, as well as their spectral composition, differs only slightly between these two regimes. Analyzing the corresponding phase portraits, however, we can say that, in the second regime, the oscillations are more nonlinear (i.e., their cycles are far more complicated in shape) than those at slow electron beam deceleration rates.

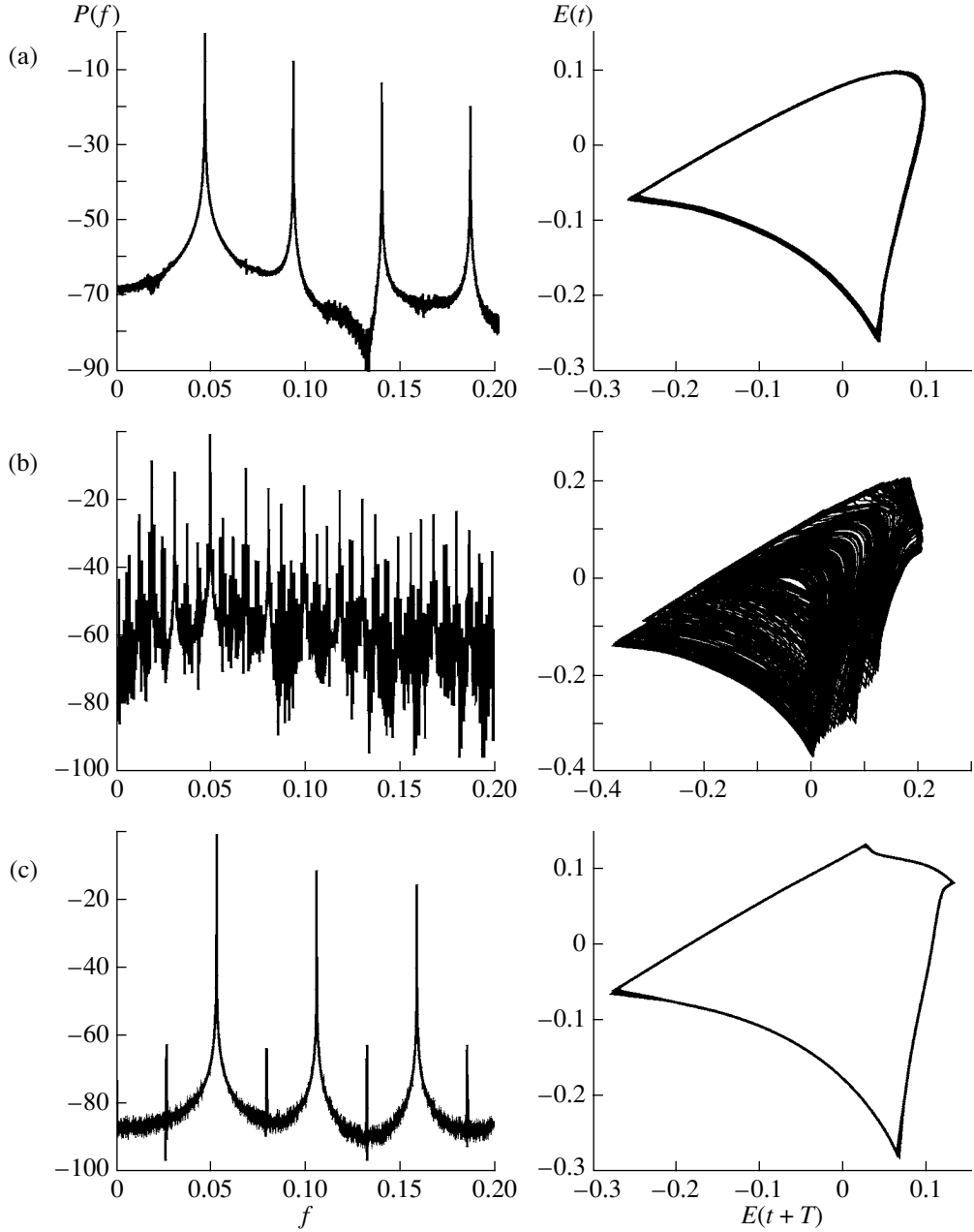

**Fig. 10.** Oscillation power spectra $P(f)$ and phase portraits of the electric field oscillations $E(t)$ at the entrance grid of the diode gap for a Pierce parameter equal to $\alpha = 0.9$ and for different values of the decelerating potential difference: $\Delta\varphi =$ (a) 0.37, (b) 0.47, and (c) 0.625.

For a larger decelerating potential difference, $\Delta\varphi \approx 0.5$, the system evolves from the regime of regular oscillations to the regime of stochastic generation (Fig. 9, domain $C$). The parameters of the electric field oscillations in the stochastic generation regime at a potential difference of $\Delta\varphi = 0.47$ are shown in Fig. 10b. We can see that, in regime $C$, the oscillation power spectrum becomes continuous, although highly irregular. The phase portrait in this case corresponds to a stochastic attractor. The system evolves from the regime of periodic oscillations to the regime of stochastic microwave generation by the VC through the regime of quasiperiodic oscillations (Fig. 9, domain $Q$). The quasiperiodic oscillation regime is characterized by oscillations at two incommensurable base frequencies, whose image in phase space is a two-dimensional torus. Accordingly, the system evolves to the stochastic generation regime in a classical way—through the disruption of quasi-periodic oscillations (and of the torus in phase space) [48].

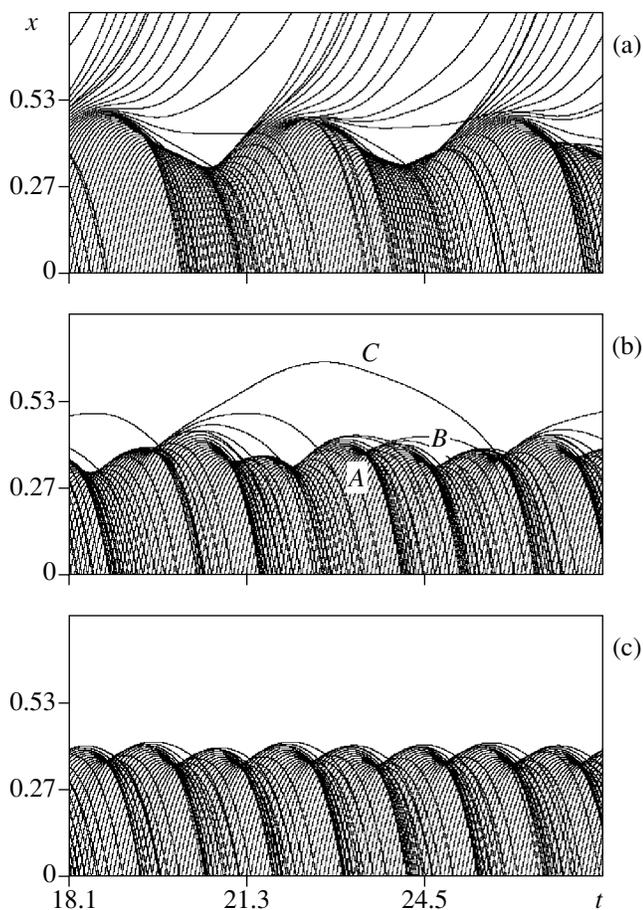

**Fig. 11.** Spatiotemporal diagrams of the dynamics of an electron beam in a diode gap for a Pierce parameter equal to $\alpha = 0.9$ and for different values of the decelerating potential difference: $\Delta\varphi =$ (a) 0.37, (b) 0.47, and (c) 0.625.

This scenario agrees well with the data of experimental investigations reported in the previous section of the paper. Thus, as the decelerating potential difference $\Delta V_{dec}$ between the grids in the model experimental device was increased, we observed the excitation of nearly regular oscillations; as the beam deceleration rate was further increased, the oscillations became more complicated. At large values of $\Delta V_{dec}$, we observed stochastic broadband microwave generation. Such behavior is qualitatively similar to that observed experimentally. A good qualitative agreement with experiment was achieved for slow deceleration rates of the electron beam—the case in which two-dimensional effects in the beam are unimportant and a one-dimensional model describes well the unsteady dynamics of an electron beam with a VC in the model experimental device.

Note that the numerically calculated dependence of the oscillation frequency in a beam with a VC on the decelerating potential is also in good qualitative agreement with that obtained experimentally (see Fig. 4). Numerical simulations, as well as experimental investigations, show that an increase in the decelerating potential difference $\Delta\varphi$ leads to an increase in the generation frequency, proportional to the plasma frequency of the beam electrons, which, in turn, increases with $\Delta\varphi_0$ (i.e., with the beam perveance).

Let us now consider the cause of such a complicated dynamics of an electron beam with an overcritical perveance in a diode gap with a decelerating field. To do this, we analyze the physical processes occurring in an electron beam with a VC in a decelerating field by examining the spatiotemporal diagrams of the electron beam and also by reconstructing the electron velocity distribution functions for the regimes of regular and of stochastic oscillations of the VC.

Figure 11 shows the spatiotemporal diagrams of the beam electrons in the dimensionless coordinates $(t, x)$ for the same values of the control parameters $\alpha$ and $\varphi_0$ as in Fig. 10. Recall that, in this section, we are working in the reditron model, in which the effect of the electrons reflected from the VC on the beam in the region ahead of the first grid, as well as the possibility that the beam electrons can repeatedly pass from the accelerating gap to the deceleration region, is ignored. Since the influence of the reflected electrons on the injected beam in the region ahead of the first (entrance) grid is ignored, the reditron model cannot describe some effects that may occur in the interaction of the reflected electrons with the injected ones, in particular, the possible additional velocity modulation of the injected electron beam. In Fig. 11, the coordinate $x = 0$ corresponds to the first grid of the diode gap and the coordinate $x = 1$ shows the position of the second grid. Each curve in the diagrams presents the trajectory of a charged macroparticle used in numerical simulations. The diagrams show the trajectories of only some macroparticles that were injected into the interaction space with the same velocity at equal time intervals. The concentration of the trajectories of the charged macroparticles corresponds to the compression—the onset of an electron structure (bunch)—in the electron beam.

An analysis of the spatiotemporal diagrams of charged macroparticles in a beam with an overcritical perveance shows that the electron velocity in the diode gap becomes lower (the slope angle of the trajectories of charged macroparticles decreases) and, at a certain spatial point in the interaction space, the electrons stop moving and turn back, so the beam passes over to a two-stream state. The turning point of the electrons can be regarded as the coordinate $x_{VC}$ of the VC. The figures show that, during the characteristic period of oscillations, the plane from which the electrons are reflected and, consequently, the position of the VC are displaced toward the injection plane. In other words, the potential barrier becomes high enough to reflect the electrons in a region far from the injection plane and then moves toward this plane, thereby oscillating within the diode gap in both time and space.

In the regime of regular oscillations at a slow beam deceleration rate (see Fig. 11a), the trajectories of the charged macroparticles in the diode gap are seen to be concentrated only in the VC region. This indicates that, in the regime of regular oscillations, only one compact dense electron bunch (structure)—a VC—forms in the system. The space charge density within the VC region is substantially higher than that in the remaining region of the interaction space. When the space charge density in the VC region becomes higher than a certain critical value, the charge is ejected from the interaction space back toward the injection plane and the space charge density in the interaction space decreases rapidly; as a result, the depth of the decelerating potential well (the VC) decreases, so the VC opens the door for the electrons, which thus easily overcome the potential barrier. The transit electrons that appear in the system are accumulated in the spatial region between the grids; as a consequence, the space charge in this region becomes higher and the depth of the potential well increases, giving rise to a VC, which begins to reflect electrons. Thereby, the process repeats itself periodically. Figure 11a shows three characteristic time periods of oscillations of the VC.

The conclusion that only one spatiotemporal structure forms in an electron beam with a VC at slow beam deceleration rates is confirmed by Fig. 12, which displays the velocity distribution functions $f(v)$ of charged macroparticles in the regime of regular oscillations, or more precisely, the electron velocity distribution functions of the electrons at the exit from the second grid (i.e., those that have passed through the VC) and the electrons at the exit from the first grid (i.e., those that have been reflected from the VC). We see from Fig. 12a that the electron velocity distribution function at the exit from the second grid in the regime of regular oscillations has only one maximum. This agrees well with the experimentally obtained single-peaked electron velocity distribution at the exit from the second grid held at a low decelerating potential at which oscillations are generated predominantly at a single frequency (cf. Fig. 6, the case $\Delta V_{dec}/V_0 = 0.1$). The velocity distribution function of the electrons reflected from the VC is more complicated in shape: in the case at hand, it has three pronounced peaks.

In the regime of stochastic oscillations (Fig. 9, domain $C$), the internal dynamics of an electron beam in a diode gap with a decelerating field is more complicated than in the previous case. Let us analyze the stochastic generation regime by referring to the spatiotemporal diagram of the electron beam (see Fig. 11b) and to the electron velocity distribution functions at the exit from the system (see Fig. 12).

From the spatiotemporal diagram we see that, during one characteristic period of oscillations in the system, two electron structures, or equivalently two VCs, form in a beam. In the diagram in Fig. 11b, they correspond to two regions of concentration of electron trajectories and are denoted by $A$ and $B$. It is clearly seen that each of the electron structures reflects some electrons. The electron structures (VCs) are strongly coupled to one another by the electrons reflected from them. The electrons reflected from one of the structures (e.g., structure $A$ in Fig. 11b) affect the potential distribution in the injection region (i.e., in the region of the first grid) and, therefore, bring additional changes in the velocities of the electrons that enter into the interaction space and form the second structure (structure $B$). In turn, the electrons reflected from the secondary structure change the "starting" conditions for the formation of the first VC in the next period of oscillations.

Such a complicated internal dynamics of the beam can be interpreted as being due to the onset of a distributed feedback in a VC-based system with an overcritical current, with the result that the system begins to generate stochastic microwave radiation (analogous results can be found in [16, 19, 20, 26]).

The complex internal dynamics of an electron beam, as well as the formation of several electron structures and the interaction between them, is also confirmed by the shapes of the velocity distribution functions of charged macroparticles in the case of a fast electron deceleration rate (see Fig. 12). In this case, the velocity distributions are more complicated in shape than those in the case of regular dynamics of an electron beam with a VC. This indicates that an electron beam in a diode gap with a decelerating field is rapidly thermalized due to the strong interaction between several electron structures near the entrance grid of the diode and, accordingly, that the nonlinear unsteady processes in the system become far more complicated [13, 49].

Note also that, in the regime of stochastic oscillations, some charged macroparticles in the system remain within the interaction space during several oscillation periods (in Fig. 11b, the trajectory of one such "long-lived" particles is denoted by $C$). Since the characteristic transit times of these particles exceed the period of oscillations of the VC, they may be thought to be responsible for the onset of the low-frequency components in the spectra of the generated oscillations when the decelerating potential of the second grid in a system with a VC is increased.

At a very large decelerating potential difference $\Delta\varphi$, the beam dynamics again is regular. The corresponding spatiotemporal diagram of an electron beam with a VC for this case is shown in Fig. 11c. We see from this diagram that the amplitude of the spatial oscillations of the VC is small and the reflected electrons are weakly modulated in velocity; as a result, at a very fast electron deceleration rate, the stochastic dynamics is suppressed. As the potential difference $\Delta\varphi$ increases fur-ther, the amplitude of the oscillations of the VC rapidly decreases and the system evolves into regime *C* (see Fig. 9), in which the VC is steady and reflects all of the beam electrons back toward the injection plane.

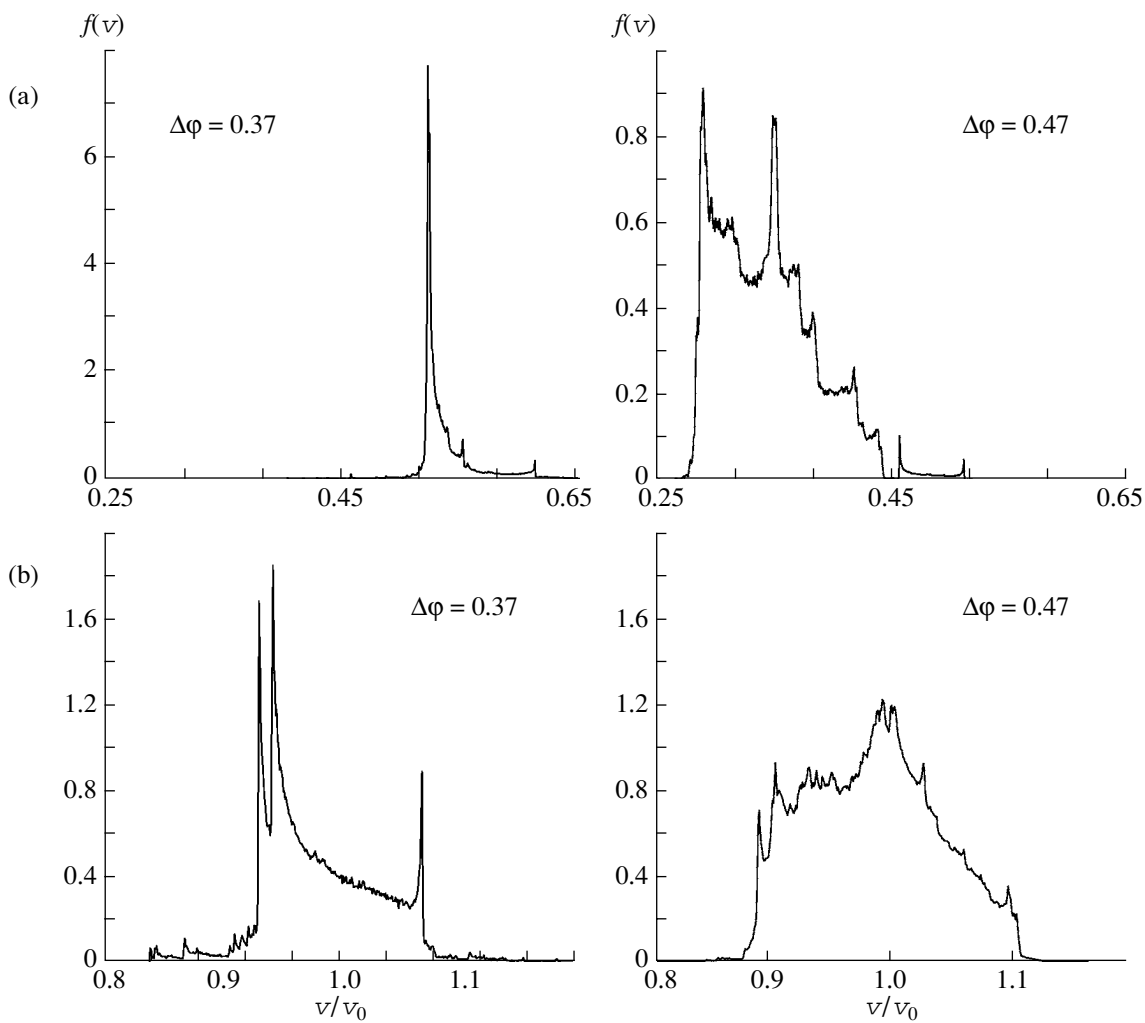

**Fig. 12.** Velocity distribution functions (a) of the electrons at the exit from the second grid (i.e., those that have passed through the VC) and (b) of the electrons at the exit from the first grid (i.e., those that have been reflected from the VC) for a Pierce parameter equal to $\alpha = 0.9$ and for $\Delta\varphi = 0.37$ and $0.47$.

## 5. DISCUSSION OF THE THEORETICAL AND EXPERIMENTAL RESULTS

Hence, our experimental investigations show that the broadband noisy oscillations observed in the drift gap with a decelerating potential are generated by a VC that forms in a nonrelativistic electron beam in a decelerating field. Numerical simulations show that the stochastic oscillations are of a dynamic nature and are governed by the processes of the formation of electron structures in a beam with an overcritical perveance and of the interaction between them. The broadening of the oscillation spectrum with increasing beam deceleration rate is attributed to the onset of secondary electron structures after the formation of the main electron structure—the VC. The secondary structures are grouped bunches of the electrons that have passed through the unsteady VC region and, correspondingly, have been modulated at a frequency close to the plasma frequency of the beam in the drift gap with a decelerating field. The appearance of the low-frequency components in the spectra of oscillations generated by a beam with an overcritical perveance can be associated with the dynamics of the electrons that remain within the interaction space during several periods of oscillations of the VC.

At a qualitative level, the results of one-dimensional numerical simulations confirm the experimental data fairly well. However, a comparison with experiment shows that the one-dimensional theory is only capable of describing regular oscillation regimes at a slow beam deceleration rate and some characteristic features of the evolution of the system into the stochastic generation regime with increasing decelerating potential. In particular, the one-dimensional theory does not provide an analytic description of the experimentally revealed effect—the change in the spectral composition of oscillations in a beam with a VC when the decelerating potential of the second grid is increased (see Fig. 3). According to [8, 37], the experimentally revealed characteristic features of the change in the spectral composition of oscillations in different cross sections of the beam with increasing decelerating potential are explained as being due to the complicated radial structure of a VC, which has the shape of a cup in the radial direction and is convex toward the injection plane. This allows us to conclude that, in order to analyze and describe the dynamics of an electron beam with a VC in a strong decelerating field (corresponding to the experimentally observed regimes of developed stochastic oscillations), it is necessary to take into account two-dimensional effects, which can be described only in terms of a two-dimensional numerical model. Such two-dimensional numerical analysis seems to be very important and it will be the subject of our further theoretical studies of electron–wave devices in which stochastic oscillations are generated by a VC that forms in an electron beam with an overcritical perveance.

It is also necessary to point out the following important effect, which was revealed by analyzing a numerical model of a diode gap with an additional deceleration of an electron beam with a VC. In [29], it was shown that, when a monoenergetic electron beam with an overcritical current (i.e., with an overcritical value of the Pierce parameter, $\alpha > \alpha_{cr} = 4/3$ [39]) is injected into a diode gap with a zero decelerating potential, it excites only regular (periodic) oscillations, no matter how much the Pierce parameter is above its critical value. With an additional nonzero decelerating potential, the dynamics of an electron beam in the diode gap becomes far more complicated: the system can even pass into the regimes of developed stochastic generation. This indicates that applying a decelerating potential to the second grid of the diode makes the regime of stochastic oscillations of the VC easier to achieve—an important point for the practical development of VC-based oscillators of stochastic microwave radiation. In our opinion, the complications of the beam dynamics in a system with a decelerating field can be attributed to the fact that the coupling between the structures forming in an electron beam with an overcritical current (perveance) becomes stronger because of the additional electron deceleration and, accordingly, because of an increasingly large number of electrons that are reflected from the secondary electron structure (the secondary VC) and return to the injection plane, thereby ensuring additional distributed feedback in the beam. On the other hand, it should be noted that our numerical simulations were carried out based on the reditron model, in which the effect of the electrons reflected from the VC on the beam in the region ahead of the first grid, as well as the possibility that the beam electrons can repeatedly pass from the gap between the electron gun and the first (entrance) grid to the deceleration region, was ignored. In Section 4, it was already mentioned that, in the reditron model, no account is taken of the effect that is exerted upon the injected beam by the reflected electrons and that may lead, e.g., to an additional velocity modulation of the injected electron beam. These factors will be the subject of our further investigations; taking them into account may be very important for achieving a better agreement between the simulation results and the experimental data (this concerns, first of all, such parameters of the spectral composition of the stochastic oscillations as the frequency bandwidth of the spectrum and its modulation depth) in the case of low decelerating potentials at the second grid and of strong guiding magnetic fields (when two-dimensional effects are insignificant).

Note in conclusion that the mechanism for generating broadband noisy oscillations that has been considered here both experimentally and theoretically can be used to create controlled microwave oscillators of high- and moderate-power broadband signals.

## ACKNOWLEDGMENTS

We are grateful to Prof. D.I. Trubetskov, corresponding member of the Russian Academy of Sciences, for his interest in this study, useful discussions, and valuable critical remarks. This work was supported in part by the Russian Foundation for Basic Research (project nos. 05-02-16273 and 05-02-16286), the Department of Atomic Science and Technology of the RF Ministry of Atomic Industry, and the US Civilian Research and Development Foundation for the Independent States of the Former Soviet Union (grant no. BRHE REC-006), the Dynasty Foundation, and the International Center for Fundamental Physics in Moscow.